\documentclass[aps, prd, twocolumn, lengthcheck, superscriptaddress, nofootinbib]{revtex4-2}%

\usepackage{amsmath,amssymb,amsfonts}
\usepackage{graphicx}
\usepackage{bm}         
\usepackage{hyperref}   
\hypersetup{breaklinks=true}   
\usepackage{physics}    
\usepackage{float, makecell, slashed}
\usepackage{subcaption}      

\usepackage[normalem]{ulem}
\usepackage{color}
\definecolor{brown}{rgb}{0.6, 0.3, 0.0}
\def\be{\begin{eqnarray}}\def\ee{\end{eqnarray}}

\begin{document}

\title{White dwarf-neutron star matter transition and the effect of light elements}

\author{Yao Ma}
\email{mayao@ucas.ac.cn}
\affiliation{School of Frontier Sciences, Nanjing University, Suzhou 215163, China}
	

\author{Yong-Liang Ma}
\email{ylma@nju.edu.cn}
\affiliation{School of Physics, Nanjing University, Nanjing 210093, China}
\affiliation{School of Frontier Sciences, Nanjing University, Suzhou 215163, China}

\author{Ruo-Xi Wu}
\email{wuruoxi@itp.ac.cn}
\affiliation{School of Fundamental Physics and Mathematical Sciences, Hangzhou Institute for Advanced Study, UCAS, Hangzhou 310024, China}
\affiliation{School of Frontier Sciences, Nanjing University, Suzhou 215163, China}

\author{Yue-Liang Wu}
\email{ylwu@ucas.ac.cn}
\affiliation{School of Fundamental Physics and Mathematical Sciences, Hangzhou Institute for Advanced Study, UCAS, Hangzhou 310024, China}
\affiliation{International Center for Theoretical Physics Asia-Pacific (ICTP-AP), UCAS, Beijing, 100190, China}
\affiliation{Institute of Theoretical Physics, Chinese Academy of Sciences, Beijing 100190, China}
\affiliation{TaiJi Laboratory for Gravitational Wave Universe (Beijing/Hangzhou), University of Chinese Academy of Sciences, Beijing 100049, China}

\date{\today} 

\begin{abstract}
White dwarfs and neutron stars are unique laboratories for dense nuclear matter physics.
We develop a single relativistic mean-field framework that treats both classes of compact star, and the transition between them, on the same footing: the nuclei of white-dwarf matter are solved self-consistently as Wigner-Seitz cells with the full electromagnetic interaction, while the same Lagrangian yields the uniform nuclear matter of the neutron-star interior.
Within this unified description we compute light-element white dwarfs seeded by $^4$He, $^{12}$C, and $^{16}$O, following each fixed-$A$ sequence along its neutronization path and connecting it to the neutron-star branch through exact Maxwell junctions, from which the corresponding mass-radius relations are derived.
The helium- and carbon-seeded white-dwarf sequences attain maximum masses of ${\sim}1.4\,M_\odot$ and ${\sim}1.0\,M_\odot$, respectively.
On the neutron-star branch, the retained light-element envelope changes the predicted radii only at the percent level---by approximately $0.2$~km at $1.4\,M_\odot$, within current observational uncertainties.
Providing a consistent zero-temperature equation of state from white-dwarf to neutron-star densities, this unified framework offers a natural starting point for studies of white-dwarf--neutron-star binary mergers, progenitor-star evolution, decihertz gravitational-wave sources, and related multimessenger phenomena.
\end{abstract}

\maketitle

\allowdisplaybreaks

\section{Introduction}
	
Compact stars---such as white dwarfs (WDs) and neutron stars (NSs)---represent the final evolution of the vast majority of stars. They serve as extraordinary laboratories for probing the behavior of matter at densities far beyond those accessible in terrestrial experiments.
On the observational front, the Gaia and Hipparcos observational data have released much information on the mass and radius of WDs~\cite{tremblay2016gaia}. The Neutron star Interior Composition Explorer (NICER) X-ray telescope has recently provided precise simultaneous mass-radius measurements of several millisecond pulsars, such as PSR~J0030+0451, PSR~J0740+6620~($\sim 2\,M_\odot$), and PSR~J0437-4715 (mass $\approx 1.42\,M_\odot$~\cite{Reardon:2024J0437, Choudhury:2024J0437, Miller:2026J0437}). Meanwhile, the binary neutron star merger GW170817 placed constraints on the tidal deformability $\tilde{\Lambda}_{1.4}$~\cite{De:2018uhw}. All these observations place stringent bounds on the dense matter equation of state (EoS) at supranuclear densities.

In the literature, after Chandrasekhar~\cite{chandrasekhar1931a, Chandrasekhar:1931ih} and Landau~\cite{Landau:1932uwv} studied WDs under the assumption of a free electron gas by ignoring the electromagnetic interaction, Feynman, Metropolis, and Teller (FMT)~\cite{Feynman:1949zz} extended the Thomas-Fermi model, a statistical model that treats the electrons surrounding a nucleus as a non-interacting uniform Fermi gas in a self-consistent electrostatic potential, to derive the EoS under extremely high pressure. Later, this approach was generalized to relativistic regimes by Rotondo et al.~\cite{Rotondo:2009cr, Rotondo:2011zz}. In addition, Salpeter and collaborators~\cite{Salpeter:1961zz, 1961ApJ...134..683H} considered the corrections to non-interacting electron gas approximation by introducing the ion lattice structure where a nucleus is surrounded by the uniformly distributed electrons with Coulomb interaction.
Nowadays, it is customary to study the internal structure of WDs with asteroseismology~\cite{1988IAUS..123..305W, 1995BaltA...4..166K} and stellar evolution by MESA~\cite{Paxton:2010ji} or LPCODE code~\cite{Althaus:2005jt, Althaus:2010eq, Renedo:2010vb, 2015MNRAS.450.3708R}, where plasma physics plays an essential role, as WDs resist gravitational collapse mainly by degenerate pressure of electron gas.

In the microscopic description of the nuclei inside WDs and NS matter, hadron dynamics plays a dominant role. For example, the Skyrme force~\cite{Agrawal:2006ea, Lesinski:2006cu, Zuo:2017njs} and low energy effective theories/models of QCD~\cite{reinhard1986nuclear,Jenkins:1990jv,Sugahara:1993wz,Lalazissis:1996rd,Todd-Rutel:2005yzo,Bernard:2007zu,Scherer:2009bt,Lattimer:2012xj,Li:2022okx,Zhang:2024sju,Ma:2023eoz,Ma:2026kun} have been verified in accordance with experiments and observed data. This approach has been widely discussed in NS physics in the literature. However, in past studies of WD matter, people typically included the effects of nuclei via point-particle approximation or semi-empirical mass formulas, without accounting for the detailed internal structure of the nucleus which is controlled by hadron dynamics.

In the previous work~\cite{Guo:2024nzi}, we treated the fine nuclear structure self-consistently within a relativistic mean-field (RMF) framework, solving for both the nuclear density profile and the surrounding electron gas on equal footing.
A key advantage of this unified framework is that it naturally enables a continuous description of the transition from WD matter to NS matter within a single consistent model, since both regimes share the same underlying Lagrangian.

The consistent treatment of WDs and NSs in a single framework is increasingly important in view of the growing observational interest in WD-NS binary systems, whose merger gravitational waves fall in the decihertz frequency band accessible to some planned observatories such as the Deci-hertz Interferometer Gravitational wave Observatory (DECIGO) and Big Bang Observer (BBO)~\cite{Kang:2024wdns}. No WD-NS system has been reported in the ground-based LIGO-Virgo-KAGRA catalogs, as expected since the WD-NS inspiral and disruption occur below the sensitive band of ground-based detectors, and whose merger products have been proposed as origins of peculiar long-duration gamma-ray bursts (e.g., GRB~211211A~\cite{Liu:2025grb} and GRB~230307A~\cite{Du:2024grb230307, Wang:2024grb230307}).

In this work, we present a unified description of WDs, NSs, and the WD-to-NS matter transition within a single RMF framework.
Specifically, we construct a composite EoS connecting WD matter to NS matter through exact Maxwell junctions, and derive the corresponding mass-radius (M-R) relations for WDs, NSs, and the transition between them, with special interest in the effect of light elements. We find that the position of the element transitions---or the type of the light element---has a non-negligible impact on the predicted M-R curves, and that a lighter-element outer envelope yields a slightly smaller NS radius.
Moreover, unlike the previous studies of WDs, our model is anchored on the microscopic degrees of freedom of compact stars---hadrons and electrons. As a result, it offers a unified description of NSs, NS crusts and WDs, benefiting future studies of phase transitions, stellar evolution, and WD-NS mergers. In addition, since our approach considered the effect of different light elements, by comparing the present results with other approaches including the corrections to the Chandrasekhar limit, one can get some insights into the effect of light elements on the formation of compact stars.

The rest of this paper is organized as follows.
Sec.~\ref{sec:eos} introduces the RMF Lagrangian and our unified description of WD and NS matter.
The neutronization process and Maxwell construction of the composite EoS are detailed in Sec.~\ref{sec:transition}.
The resulting M-R relations for the WD-to-NS transition, WDs, and NSs are presented in Sec.~\ref{sec:struc}.
Finally, discussion and outlook are given in Sec.~\ref{sec:concl}.

\section{EoS construction}
 \label{sec:eos}

To investigate the main features of both WD matter and NS matter, we consider a typical Walecka model with electromagnetic interaction~\cite{Sugahara:1993wz,Shen:1998gq} whose Lagrangian is given as
\be
\mathcal{L} 
& = & \bar{\Psi}\left(i \slashed{\text D} - m_N\right)\Psi + \bar{\psi}\left(i \slashed{\text D} -m_e\right)\psi \nonumber\\
& &{} + \frac{1}{2}(\partial_{\mu}\sigma\partial^{\mu}\sigma - m_{\sigma}^2\sigma^2) -\frac{1}{3}g_2\sigma^3-\frac{1}{4}g_3\sigma^4 \nonumber\\
& &{} - \frac{1}{4}\Omega_{\mu\nu}\Omega^{\mu\nu} +\frac{1}{2}m_{\omega}^2\omega_{\mu}\omega^{\mu} + \frac{1}{4}c_3\left(\omega_{\mu}\omega^{\mu}\right)^2 \nonumber\\
& &{} - \frac{1}{4}\vec{\bm P}_{\mu\nu}\cdot\vec{\bm P}^{\mu\nu} +\frac{1}{2}m_{\rho}^2\vec{\rho}_{\mu}\cdot\vec{\rho}^{\mu} -\frac{1}{4}F_{\mu\nu}F^{\mu\nu} \nonumber \\
& &{} + \bar{\Psi}\left(-g_{\sigma}\sigma-g_{\omega}\slashed{\omega}-g_{\rho}\vec{\slashed{\rho}}\right)\Psi\ , 
\label{eq:LagTM2}
\ee
where $\Psi=\left(\begin{array}{l}p \\ n\end{array}\right)$ is the nucleon iso-doublet with $m_N = \mathrm{diag}(m_p, m_n)$ (the NN1 set uses $m_n \neq m_p$; see Sec.~\ref{sec:struc}), $\psi$ is the electron field,
$A_\mu$ is the electromagnetic field.
$\sigma, \omega_{\mu}$ and $\vec{\rho}_{\mu}=\rho_{\mu}^i\tau^i$ (with $\tau_i$ being the Pauli matrix) are isoscalar-scalar, isoscalar-vector and isovector-vector meson fields, respectively.
The pseudoscalar mesons $\pi$ are neglected since they vanish in the RMF approximation.
$\vec{\bm P}_{\mu\nu} = \text D_{\mu}\vec{\rho}_{\nu}-\text D_{\nu}\vec{\rho}_{\mu} - 2ig_\rho\vec{\rho}_{\mu}\times\vec{\rho}_{\nu}$ is the field-strength tensor of $\rho$ meson fields.
The covariant derivatives are defined as
\be
\text D_{\mu}\Psi & = & \left(\partial_{\mu}+iA_{\mu}Q\right)\Psi
\ , \nonumber\\
\text D_{\mu}\psi & = & \left(\partial_{\mu}-ieA_{\mu}\right)\psi\ ,\nonumber\\
\text D_{\mu}\vec{\rho}_{\nu} & = & \partial_{\mu}\vec{\rho}_{\nu}+iA_{\mu}\left[Q,\vec{\rho}_{\nu}\right]\ ,
\ee
with the charge matrix $Q = e \left(1+\tau_3\right)/2$.

In the construction of WD matter, the matter is regarded as a composite system made of identically spherical Wigner-Seitz cells (atom-like units), each consisting of a nucleus at the center surrounded by a homogeneous electron gas with the cell radius determined by baryon number density~\cite{Guo:2024nzi}.
The nucleus is described within the RMF framework using the Walecka-type Lagrangian~\eqref{eq:LagTM2}, and the full electromagnetic interaction is solved self-consistently via Maxwell equations.
Both the WD shape and constituent Wigner-Seitz cell are assumed to be spherically symmetric.

The high-density branch is uniform nuclear matter (NM) computed with the same Walecka-type Lagrangian following the standard RMF procedure~\cite{Guo:2024nzi}. Consistently with the fully neutronized endpoints of the fixed-$A$ sequences, it is evaluated as pure neutron matter, and the small beta-equilibrium proton fraction of $npe\mu$ matter is neglected. This stiffens the core EoS at the level of a few tenths of MeV per-baryon around saturation density, and a self-consistent $npe\mu$ rerun of the NS branch is left to a forthcoming revision.
In the present calculation this uniform NM branch is joined to the Wigner-Seitz branche directly, the inhomogeneous inner crust---nuclear clusters immersed in a dripped-neutron gas, with possible pasta phases---is not modeled. This issue will be clarifies in the future.

The transition from WD matter to NS matter is governed by the beta-equilibrium condition
\begin{equation}
  \mu_n = \mu_p + \mu_e \;, \label{eq:beta}
\end{equation}
where $\mu_n$, $\mu_p$, and $\mu_e$ are the chemical potentials of neutrons, protons, and electrons, respectively.
This condition determines the preferred sub-state at each pressure by minimizing the Gibbs free energy per-baryon, which defines the neutronization path. Away from a transition pressure, each macroscopic phase is represented by a single species of locally neutral Wigner-Seitz cell of sub-state $(N_n, N_p)$, with $N_{n (p)}$ referring to the number of neutrons (protons) in the cell. At a first-order transition the two phases coexist with the same pressure and baryon chemical potential but different baryon densities. As the pressure increases, neutronization proceeds via $p + e^- \to n + \nu_e$, converting proton-rich WD matter step by step into neutron-rich NS matter.
The energetically stable ground state is identified at each pressure as the sub-state with the lowest Gibbs free energy per-baryon,
\begin{equation}
  \mu_B \equiv \frac{\varepsilon + p}{n_B} \;, \label{eq:gibbs}
\end{equation}
where $\varepsilon$ is the total energy density (nucleons and electrons, with a common rest-mass convention) and $n_B$ the baryon number density. At zero temperature $\mu_B$ coincides with the baryon chemical potential, so this criterion is equivalent to selecting, at fixed $\mu_B$, the sub-state with the maximum pressure.
The individual sub-state EoSs are then joined by the Maxwell construction~\cite{Glendenning:2000} carried out in the $p$--$\mu_B$ plane: two sub-states coexist at the point where their $p(\mu_B)$ curves cross, which enforces simultaneously the mechanical ($p_A = p_B$) and chemical ($\mu_{B,A} = \mu_{B,B}$) equilibrium conditions, as detailed in the following.

\section{Transition from white dwarf matter to neutron star matter}
\label{sec:transition}

For a WD made of elemental composition $^A$X, the total baryon number per-Wigner-Seitz cell, $N = N_n + N_p = A$, is conserved throughout neutronization. The construction below therefore yields the conditional equilibrium within this restricted fixed-$A$ family of locally neutral cell branches. Note that it is not the unrestricted cold-catalyzed ground state of Baym-Pethick-Sutherland type~\cite{Baym:1971pw, Haensel:1994}, in which both $A$ and $Z$ are optimized, but rather reflects the composition inherited from the progenitor WD.
Successive sub-states differ by replacing one proton with a neutron, i.e., $(N_n, N_p) \to (N_n+1, N_p-1)$, driven by the beta-equilibrium condition Eq.~\eqref{eq:beta}. The stable ground state at each pressure is the sub-state $(N_n, N_p)$ with the lowest Gibbs free energy per-baryon $\mu_B$ of Eq.~\eqref{eq:gibbs}. Each branch is represented by its curve $p(\mu_B)$, and the ground-state EoS is the upper envelope of these curves.

\begin{figure*}[htb]
    \centering
    \begin{subfigure}[t]{0.48\textwidth}
        \centering
        \includegraphics[width=\linewidth]{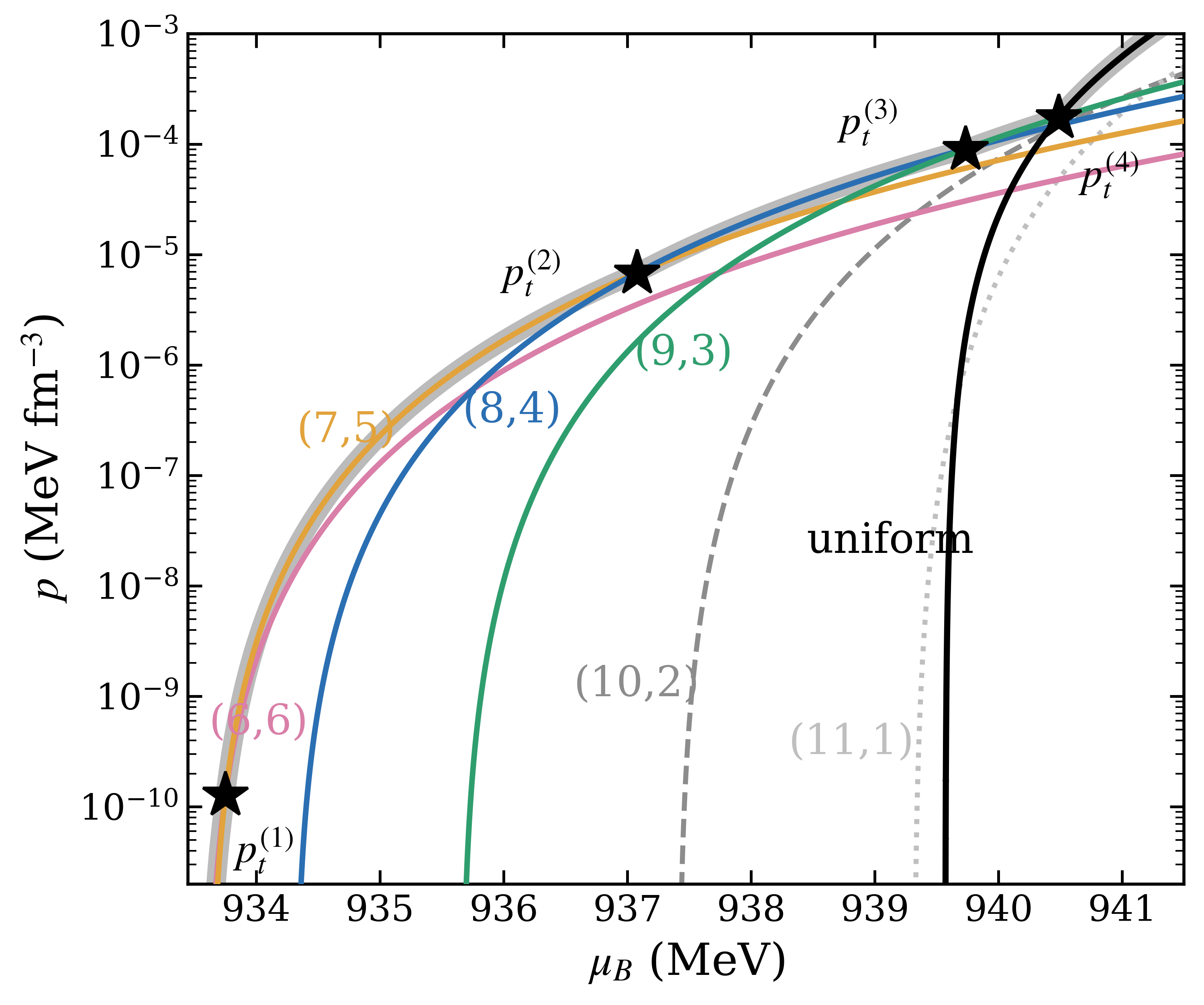}
        \caption{Sub-state branches of $^{12}$C in the $p$--$\mu_B$ plane. At each pressure the stable sub-state minimizes the Gibbs free energy per-baryon (thick gray envelope). Stars mark the equilibrium crossings.}
        \label{fig:maxwell_pmu}
    \end{subfigure}\hfill
    \begin{subfigure}[t]{0.48\textwidth}
        \centering
        \includegraphics[width=\linewidth]{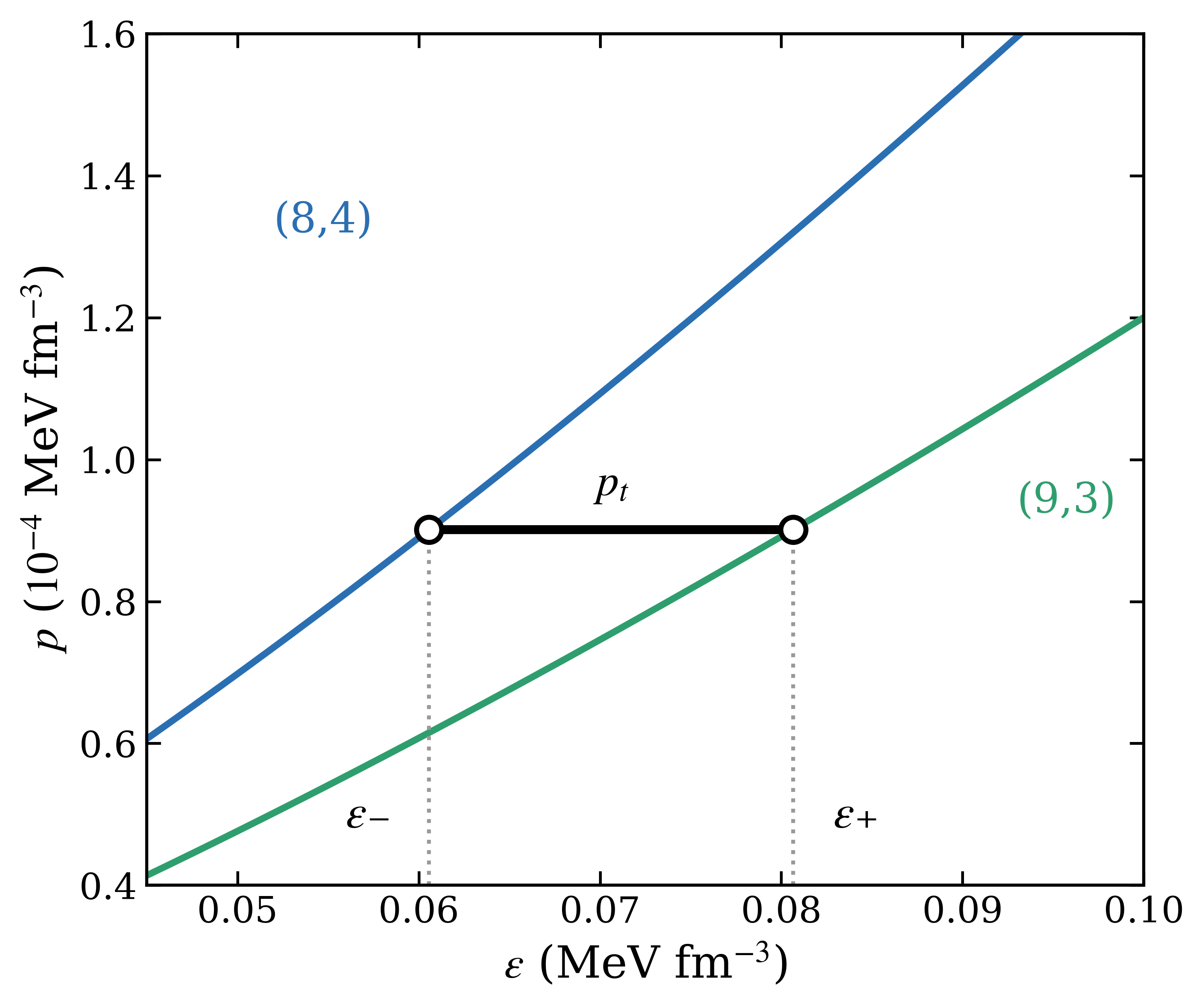}
        \caption{The unique Maxwell point for the $(8,4)\to(9,3)$ transition in the $p$--$\varepsilon$ plane. The constant-pressure segment at $p_t$ connects $(\varepsilon_-,p_t)$ to $(\varepsilon_+,p_t)$, where both mechanical and chemical equilibrium hold.}
        \label{fig:maxwell_pe}
    \end{subfigure}
    \caption{Maxwell construction for the neutronization of $^{12}$C.}
    \label{fig:rho-e_crossover}
\end{figure*}

Taking $^{12}$C as an example ($N = 12$), the first transitions proceed in natural order: $(6,6) \to (7,5) \to (8,4) \to (9,3)$, with the unique equilibrium pressures $p_t \simeq 1.3\times10^{-10}$, $6.8\times10^{-6}$, and $9.0\times10^{-5}~\mathrm{MeV/fm^3}$ and density jumps of $+20\%$, $+25\%$, and $+33\%$, respectively.
Beyond $(9,3)$, however, the $p(\mu_B)$ curve of uniform NM crosses that of the $(9,3)$ branch at a lower chemical potential than any of the more neutron-rich sub-states.
Consequently, within the restricted fixed-$A$ candidate set the minimum-Gibbs envelope skips the intermediate sub-states $(10,2)$ and $(11,1)$ and next selects the extrapolated uniform-matter branch at $p_t \simeq 1.7\times10^{-4}~\mathrm{MeV/fm^3}$ with a density jump of a factor ${\sim}3.8$, as illustrated in Fig.~\ref{fig:rho-e_crossover}(a).
The complete ground-state path for $^{12}$C is therefore $(6,6) \to (7,5) \to (8,4) \to (9,3) \to$ uniform matter.
Within the branches included here, the $(10,2)$ and $(11,1)$ sub-states never lie on the minimum-Gibbs energy envelope. In vacuum, the corresponding nuclei $^{12}$He and $^{12}$H have never been observed experimentally~\cite{Kelley:2017A12, Kondev:2021NUBASE}, and even $^{12}$Li---whose $(9,3)$ branch does lie on the envelope---is neutron-unbound~\cite{Aksyutina:2008, Hall:2010Li12}. The ordering found here therefore reflects the dense stellar environment, where the competitiveness of a branch is determined by the in-medium Gibbs energy per-baryon rather than by vacuum stability. Each transition between sub-states is of first order, with a finite density jump at the equilibrium pressure determined in the $p$--$\mu_B$ plane.

Because the transition between two sub-states is of first order, pointwise selection of the lower-energy homogeneous branch at fixed density produces a non-convex $\varepsilon(n)$ and hence a spurious region where the pressure decreases with increasing density (cf. Fig.~\ref{fig:rho-e_crossover}). The physical EoS is the convex hull of $\varepsilon(n)$, obtained by a common-tangent construction, which is equivalent to taking the crossing of the $p(\mu_B)$ curves~\cite{Glendenning:2000, Baym:1971pw}.
We therefore construct the connection directly in the $p$--$\mu_B$ plane. At each change of the global-envelope winner, for the outgoing branch A and incoming branch B we solve
\begin{equation}
  p_A(\mu_B^{t}) = p_B(\mu_B^{t}) \equiv p_t \,, \label{eq:pn}
\end{equation}
for the crossing point $(\mu_B^{t}, p_t)$ of the two $p(\mu_B)$ curves. At this single point both the mechanical equilibrium ($p_A = p_B$) and the chemical equilibrium ($\mu_{B,A} = \mu_{B,B}$) conditions are satisfied simultaneously, so the transition pressure is uniquely determined and carries no free parameter.
At $p_t$ the baryon density jumps from $n_-$ (on branch A) to $n_+$ (on branch B); in the $p$--$\varepsilon$ EoS the transition appears as a constant-pressure segment connecting $(\varepsilon_-, p_t)$ to $(\varepsilon_+, p_t)$, along which the convexified EoS obeys the lever rule $n_B = (1-x)\,n_- + x\,n_+$ and $\varepsilon = (1-x)\,\varepsilon_- + x\,\varepsilon_+$ with $x \in [0,1]$ at constant $p_t$ and $\mu_B^t$. In a hydrostatic star this Maxwell segment is realized as a sharp interface at a single radius rather than an extended constant-pressure layer. A multicomponent lattice can smooth the ideal discontinuity only over a very narrow pressure interval, ${\sim}10^{-4}\,p_t$~\cite{Haensel:1994}.
This construction is illustrated in Fig.~\ref{fig:rho-e_crossover}(b).

The complete set of junctions for the three seed compositions (and for the TM2 comparison sequence) is listed in Table~\ref{tab:junctions}.

\begin{table*}[htb]
    \caption{All Maxwell junctions of the connected EoSs (NN1 parameter set; the last two rows give the $^4$He-seeded sequence with TM2 shown in Fig.~\ref{fig:mr_tm}). For each first-order transition between the outgoing and incoming sub-states we list the transition pressure $p_t$, the baryon chemical potential $\mu_B^t$, and the coexistence baryon and energy densities. The final junction of each sequence is a proxy connection to the extrapolated uniform branch (see text).}
    \label{tab:junctions}
    \begin{tabular}{llcccccc}
        \hline\hline
        Sequence & Transition & $p_t$ (MeV\,fm$^{-3}$) & $\mu_B^t$ (MeV) & $n_-$ (fm$^{-3}$) & $n_+$ (fm$^{-3}$) & $\varepsilon_-$ (MeV\,fm$^{-3}$) & $\varepsilon_+$ (MeV\,fm$^{-3}$) \\
        \hline
        $^4$He   & $(2,2)\to(3,1)$            & $2.89\times10^{-6}$  & $939.38$ & $3.26\times10^{-6}$ & $6.51\times10^{-6}$ & $3.06\times10^{-3}$ & $6.12\times10^{-3}$ \\
                 & $(3,1)\to$ uniform          & $7.42\times10^{-6}$  & $939.85$ & $1.32\times10^{-5}$ & $6.88\times10^{-5}$ & $1.24\times10^{-2}$ & $6.46\times10^{-2}$ \\
        \hline
        $^{12}$C & $(6,6)\to(7,5)$            & $1.28\times10^{-10}$ & $933.75$ & $2.34\times10^{-9}$ & $2.80\times10^{-9}$ & $2.19\times10^{-6}$ & $2.62\times10^{-6}$ \\
                 & $(7,5)\to(8,4)$            & $6.82\times10^{-6}$  & $937.08$ & $7.48\times10^{-6}$ & $9.33\times10^{-6}$ & $7.00\times10^{-3}$ & $8.74\times10^{-3}$ \\
                 & $(8,4)\to(9,3)$            & $9.01\times10^{-5}$  & $939.73$ & $6.46\times10^{-5}$ & $8.59\times10^{-5}$ & $6.06\times10^{-2}$ & $8.07\times10^{-2}$ \\
                 & $(9,3)\to$ uniform          & $1.75\times10^{-4}$  & $940.49$ & $1.41\times10^{-4}$ & $5.31\times10^{-4}$ & $1.33\times10^{-1}$ & $4.99\times10^{-1}$ \\
        \hline
        $^{16}$O & $(9,7)\to(10,6)$           & $9.21\times10^{-7}$  & $935.20$ & $1.60\times10^{-6}$ & $1.86\times10^{-6}$ & $1.49\times10^{-3}$ & $1.74\times10^{-3}$ \\
                 & $(10,6)\to(11,5)$          & $2.57\times10^{-5}$  & $937.78$ & $2.25\times10^{-5}$ & $2.69\times10^{-5}$ & $2.11\times10^{-2}$ & $2.52\times10^{-2}$ \\
                 & $(11,5)\to(12,4)$          & $1.42\times10^{-4}$  & $939.82$ & $9.71\times10^{-5}$ & $1.21\times10^{-4}$ & $9.11\times10^{-2}$ & $1.14\times10^{-1}$ \\
                 & $(12,4)\to$ uniform         & $2.70\times10^{-4}$  & $940.64$ & $1.96\times10^{-4}$ & $7.15\times10^{-4}$ & $1.84\times10^{-1}$ & $6.73\times10^{-1}$ \\
        \hline
        $^4$He (TM2) & $(2,2)\to(3,1)$        & $4.09\times10^{-6}$  & $937.93$ & $4.23\times10^{-6}$ & $8.45\times10^{-6}$ & $3.97\times10^{-3}$ & $7.92\times10^{-3}$ \\
                 & $(3,1)\to$ uniform          & $8.12\times10^{-6}$  & $938.29$ & $1.41\times10^{-5}$ & $7.20\times10^{-5}$ & $1.32\times10^{-2}$ & $6.75\times10^{-2}$ \\
        \hline\hline
    \end{tabular}
\end{table*}

The location of the final junction onto the uniform branch deserves an explicit caveat. For all three compositions it falls near the neutron-drip region ($\rho \sim 10^{10}$--$10^{12}~\mathrm{g\,cm^{-3}}$), i.e., far below the crust-core transition of standard unified EoSs ($n_{cc} \simeq 0.076~\mathrm{fm^{-3}}$, $p_{cc} \simeq 0.34~\mathrm{MeV\,fm^{-3}}$~\cite{Douchin:2001sv}). Physically, this crossing marks the point where the restricted bound-cell branches lose to the extrapolated uniform branch because the present model lacks the lower-free-energy intermediate configurations---nuclear clusters coexisting with a dripped-neutron gas and, at higher densities, pasta phases~\cite{Chamel:2008ca}. The junction should therefore be regarded as a proxy for the onset of this missing inner-crust physics rather than as a genuine transition to uniform matter; a self-consistent inner-crust calculation is deferred to future work.

\begin{figure*}[htb]
    \centering
    \begin{subfigure}[t]{0.48\textwidth}
        \centering
        \includegraphics[width=\linewidth]{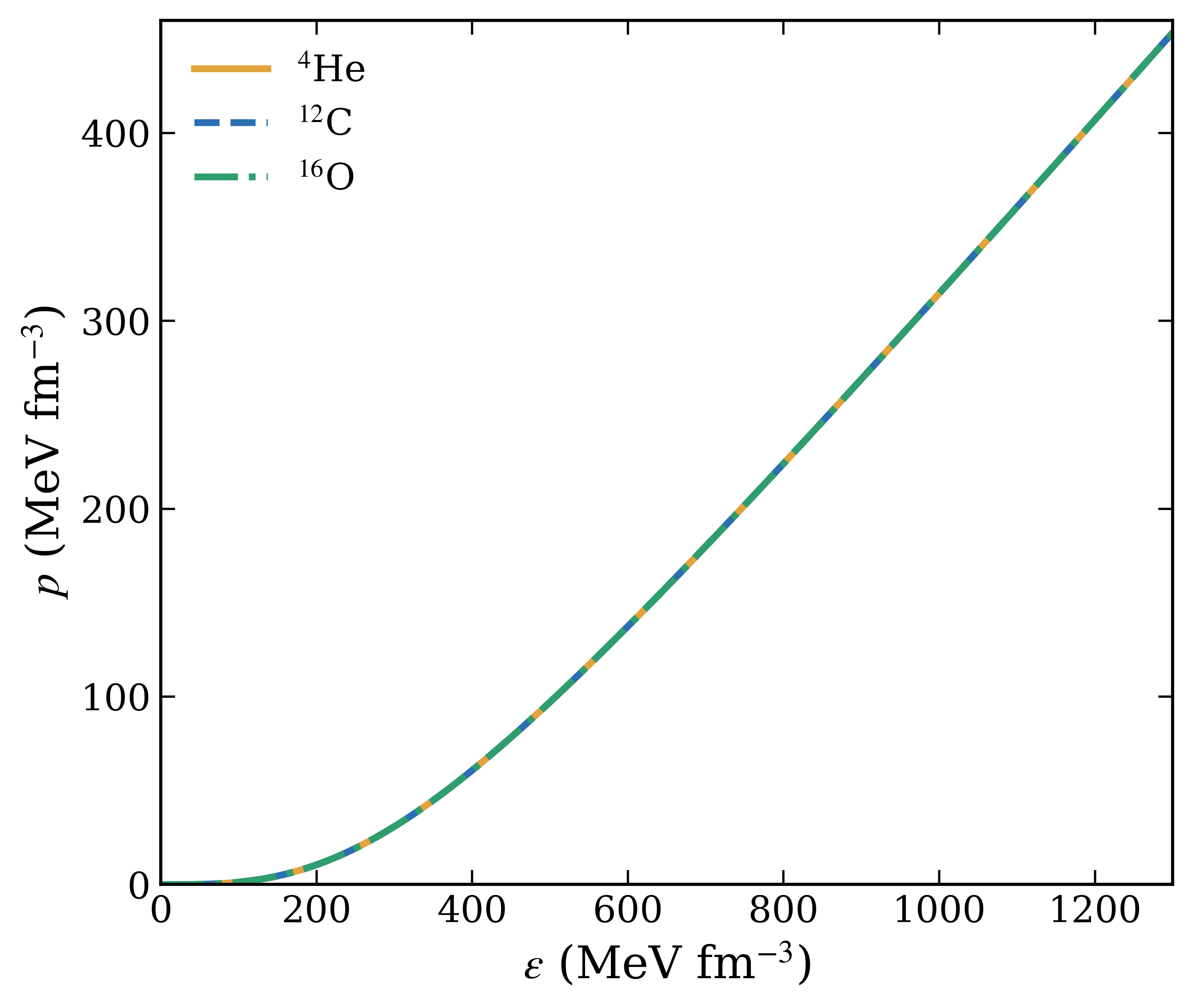}
        \caption{Global $p$--$\varepsilon$ relation for the $^4$He-, $^{12}$C-, and $^{16}$O-seeded sequences, which share the uniform-matter branch at high density.}
        \label{fig:eos_pmu_global}
    \end{subfigure}\hfill
    \begin{subfigure}[t]{0.48\textwidth}
        \centering
        \includegraphics[width=\linewidth]{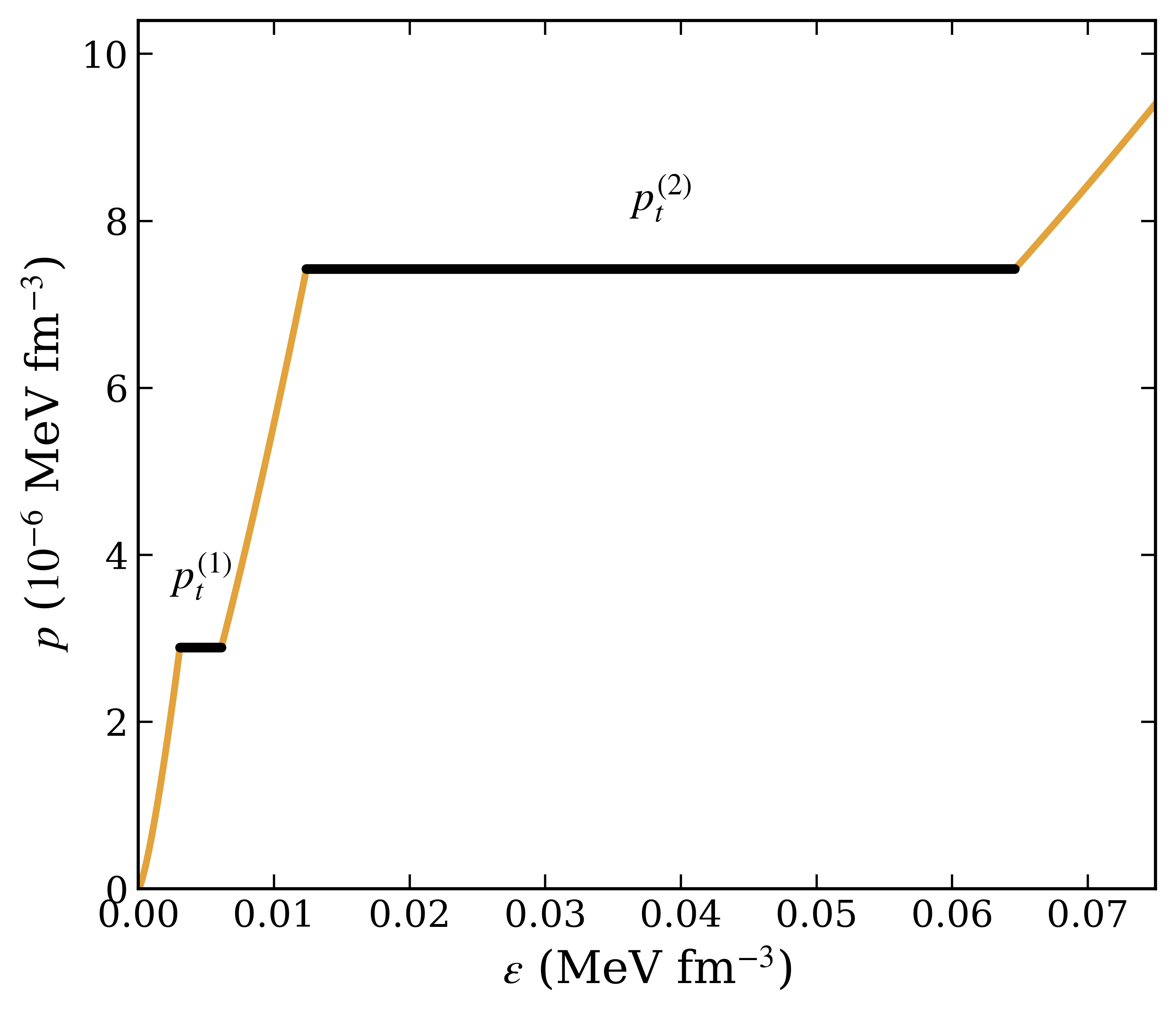}
        \caption{Enlargement of the transition region for the $^4$He-seeded sequence; black horizontal segments mark the constant-pressure Maxwell jumps $p_t^{(i)}$.}
        \label{fig:eos_pmu_he}
    \end{subfigure}\\[2pt]
    \begin{subfigure}[t]{0.48\textwidth}
        \centering
        \includegraphics[width=\linewidth]{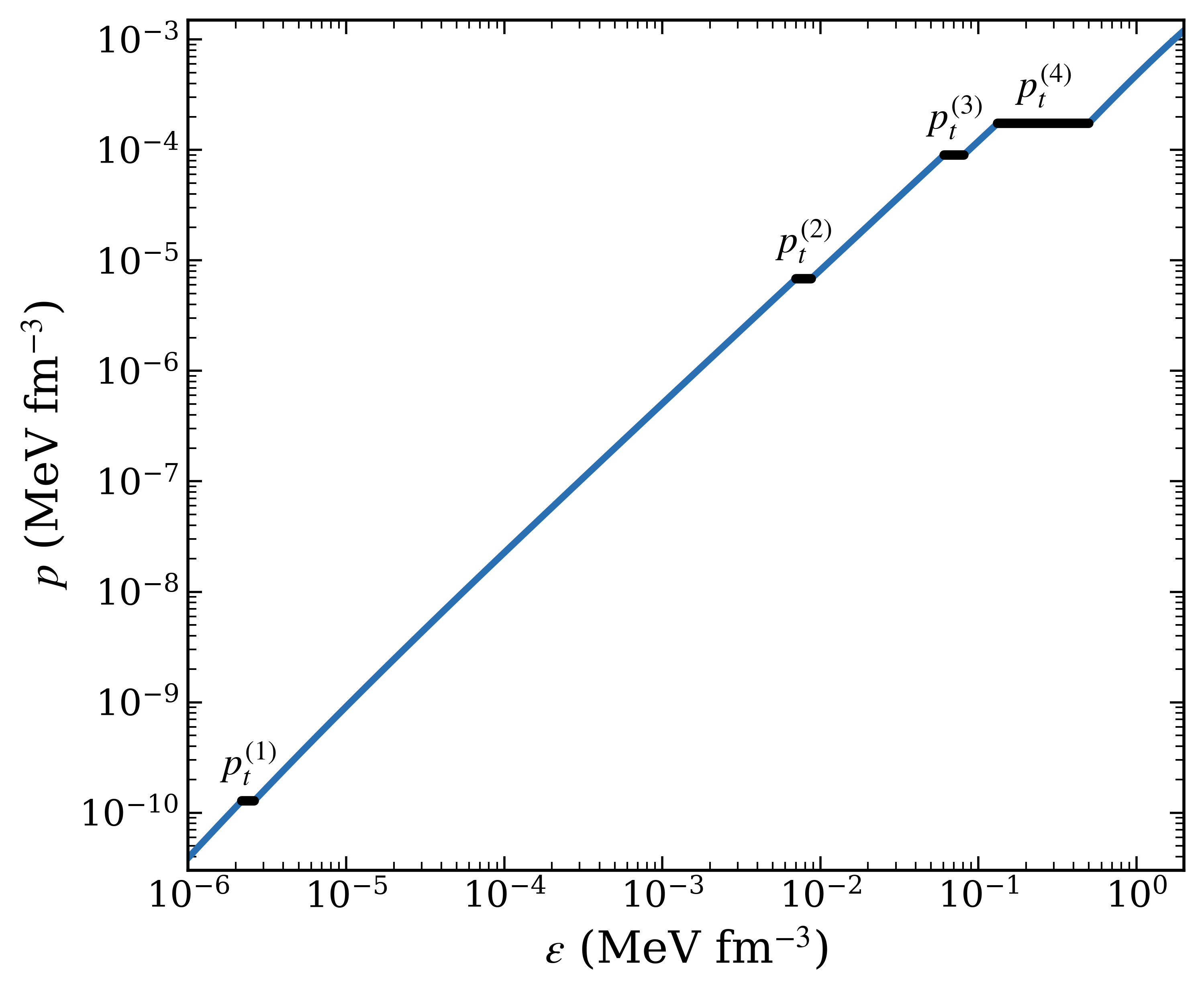}
        \caption{Same as (b) for the $^{12}$C-seeded sequence.}
        \label{fig:eos_pmu_c12}
    \end{subfigure}\hfill
    \begin{subfigure}[t]{0.48\textwidth}
        \centering
        \includegraphics[width=\linewidth]{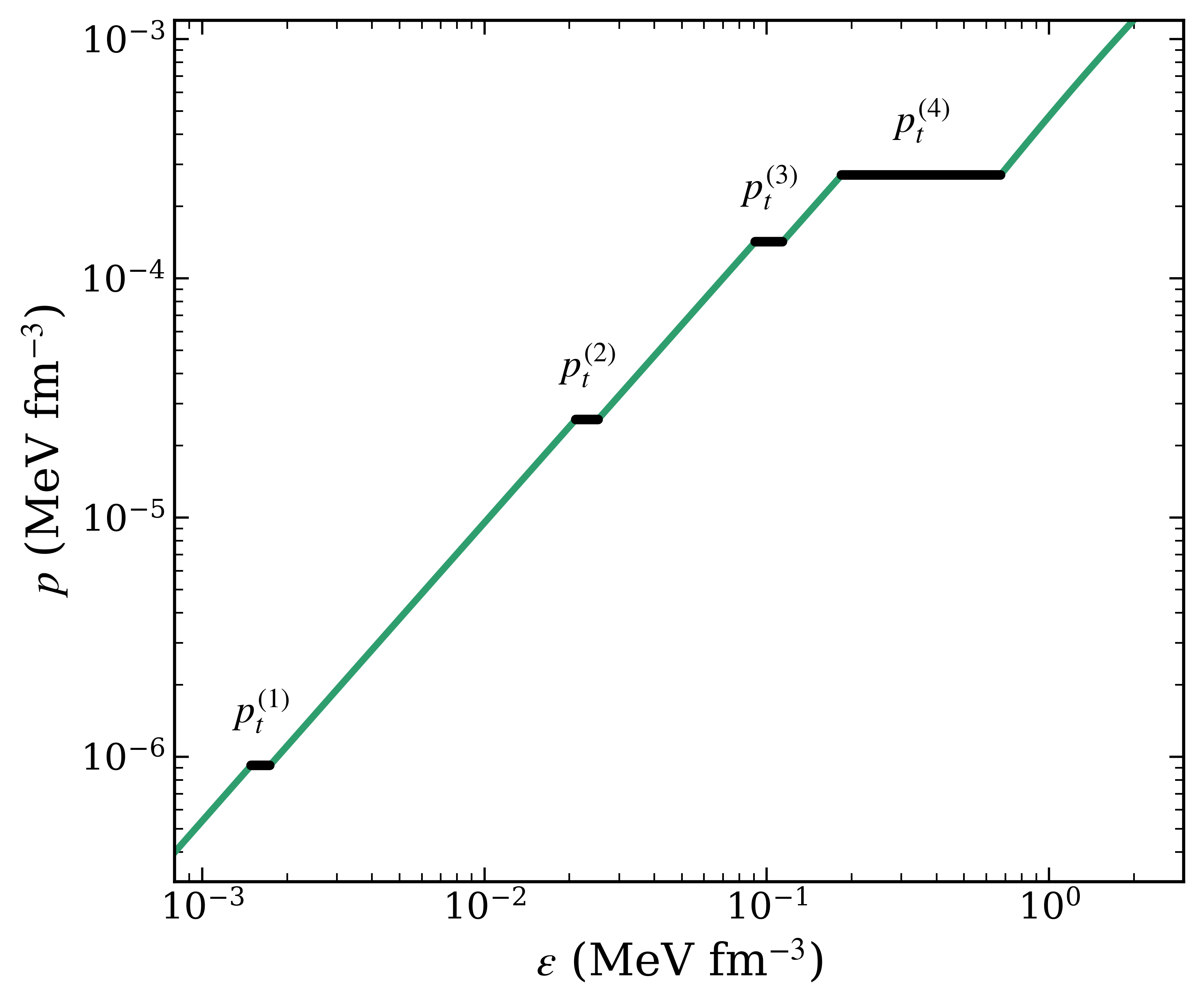}
        \caption{Same as (b) for the $^{16}$O-seeded sequence.}
        \label{fig:eos_pmu_o16}
    \end{subfigure}
    \caption{Connected EoSs with every junction at its unique equilibrium pressure.}
    \label{fig:eos_pmu}
\end{figure*}

The resulting pressure-energy density EoSs for the $^4$He-, $^{12}$C-, and (nominal) $^{16}$O-seeded restricted sequences are shown in Fig.~\ref{fig:eos_pmu} by using the NN1 parameter set discussed in the following.
Each figure displays the global EoS and a local enlargement of the Maxwell-construction region. 

\section{Structure of compact stars}
 \label{sec:struc}
Using the EoSs with the Maxwell ansatz constructed above, we next solve the Tolman-Oppenheimer-Volkoff (TOV) equations to obtain the M-R relations of compact objects across the full WD-to-NS density region for the purpose of investigating the transition of nuclei on the compact star structure. The stellar surface is located where the pressure falls to $p_{\rm surf} = 10^{-17}~\mathrm{MeV\,fm^{-3}}$. We checked that the WD radii are sensitive to the choice of stellar surface pressure---truncating the integration at $p \gtrsim 10^{-12}~\mathrm{MeV\,fm^{-3}}$ ($\rho \sim 10^{5}~\mathrm{g\,cm^{-3}}$) would underestimate the WD radii by $30$--$40\%$ at $0.5$--$0.8\,M_\odot$---whereas the NS radii are insensitive to it. The WD radii are converged to better than $0.5\%$ for $p_{\rm surf} \lesssim 10^{-16}~\mathrm{MeV\,fm^{-3}}$ so that our present choice $p_{\rm surf} = 10^{-17}~\mathrm{MeV\,fm^{-3}}$ is small enough. We present in this section the numerical results for the finite-nuclei and compact star properties. The calculations were performed by using the framework developed in our recent work~\cite{Ma:2026jff}.

\subsection{Model parameters and validation against finite nuclei}
\label{subsec:valid}

We typically employ three RMF parameter sets to validate the NM input of our model. The first two sets are the parameters estimated by using the neural network (NN) with the constraints from nuclear physics and NS measurements~\cite{Guo:2023mhf} but taking $m_n \neq m_p$ (denoted as NN1) and $m_n = m_p$ (denoted as NN2).
The third is the standard TM2 parameter set (labeled as TM2)~\cite{Sugahara:1993wz}.
The NM properties around saturation density from the NN approach---including saturation density, binding energy per nucleon, incompressibility, symmetry energy, and its slope---were tabulated in Ref.~\cite{Guo:2024nzi}. The explicit coupling constants, meson masses and nonlinear couplings of the three sets are those of Refs.~\cite{Guo:2023mhf} (NN1 and NN2) and~\cite{Sugahara:1993wz} (TM2), and are provided, together with the finite-cell field equations and boundary conditions, in the machine-readable material accompanying this work.
Table~\ref{tab:nucpro} compares the root-mean-square matter radius $\sqrt{\langle r^2\rangle}$ and binding energy per nucleon B.E.\ predicted by each parameter set against experimental values for the three $A=16$ isobars $^{16}$O, $^{16}$N, and $^{16}$C.
The NN parameters reproduce the experimental B.E.\ of the neutron-rich species $^{16}$N and $^{16}$C significantly better than TM2 which overestimates their binding energies by ${\sim}3$--$4$~MeV.
Enforcing $m_n = m_p$ brings the B.E.\ of $^{16}$O closer to experiment, but at the cost of degraded agreement for $^{16}$N and $^{16}$C, confirming that the physical neutron--proton mass difference is essential for correctly describing neutron-rich NM.
Charge radii are largely insensitive to the three choices.

\begin{table}[htbp]
    \caption{Nuclear properties of $A=16$ isobars under three RMF parameter sets.
    The radius $\sqrt{\langle r^2\rangle}$ is computed as
    $\langle r^2\rangle=\int_0^{R_0} R^2\rho(R)\,4\pi R^2\,\mathrm{d}R \big/ \int_0^{R_0} \rho(R)\,4\pi R^2\,\mathrm{d}R$,
    where $\rho(R)$ is the radial nucleon density and $R_0$ is the cell radius.
    The binding energy per-nucleon $\mathrm{B.E.} = [N m_n + Z m_p - E_i^{\rm nuc}]/(N+Z)$, where $E_i^{\rm nuc}$ is the nuclear energy per-Wigner-Seitz cell (electron contributions and rest masses treated consistently across all entries) and $N$, $Z$ are the neutron and proton numbers.
    Experimental values from Refs.~\cite{Kanungo:2016tmz,Kaur:2022yoh,NNDCNuDat}.
    $\sqrt{\langle r^2\rangle}$ is in unit of fm and B.E. in unit of MeV.}
    \label{tab:nucpro}
    \centering
    \small
    \setlength{\tabcolsep}{4pt}
    \renewcommand{\arraystretch}{1.2}
    \begin{tabular}{ccccc}
        \hline
        & Parameters & $^{16}$O & $^{16}$N & $^{16}$C \\
        \hline
        $\sqrt{\langle r^2\rangle}$ (Exp.) & & $2.57(2)$ & -- & $2.74(3)$ \\
        \hline
        $\sqrt{\langle r^2\rangle}$ (The.) & \makecell{NN1\\NN2\\TM2}
            & \makecell{2.49\\2.49\\2.69}
            & \makecell{2.50\\2.50\\2.49}
            & \makecell{2.53\\2.53\\2.52} \\
        \hline
        B.E.\ (Exp.) & & $7.98$ & $7.37$ & $6.92$ \\
        \hline
        B.E.\ (The.) & \makecell{NN1\\NN2\\TM2}
            & \makecell{7.12\\8.01\\7.86}
            & \makecell{7.02\\7.99\\11.00}
            & \makecell{6.52\\7.58\\10.37} \\
        \hline
    \end{tabular}
\end{table}

\begin{figure*}[htb]
    \centering
    \begin{subfigure}[t]{0.48\textwidth}
        \centering
        \includegraphics[width=\linewidth]{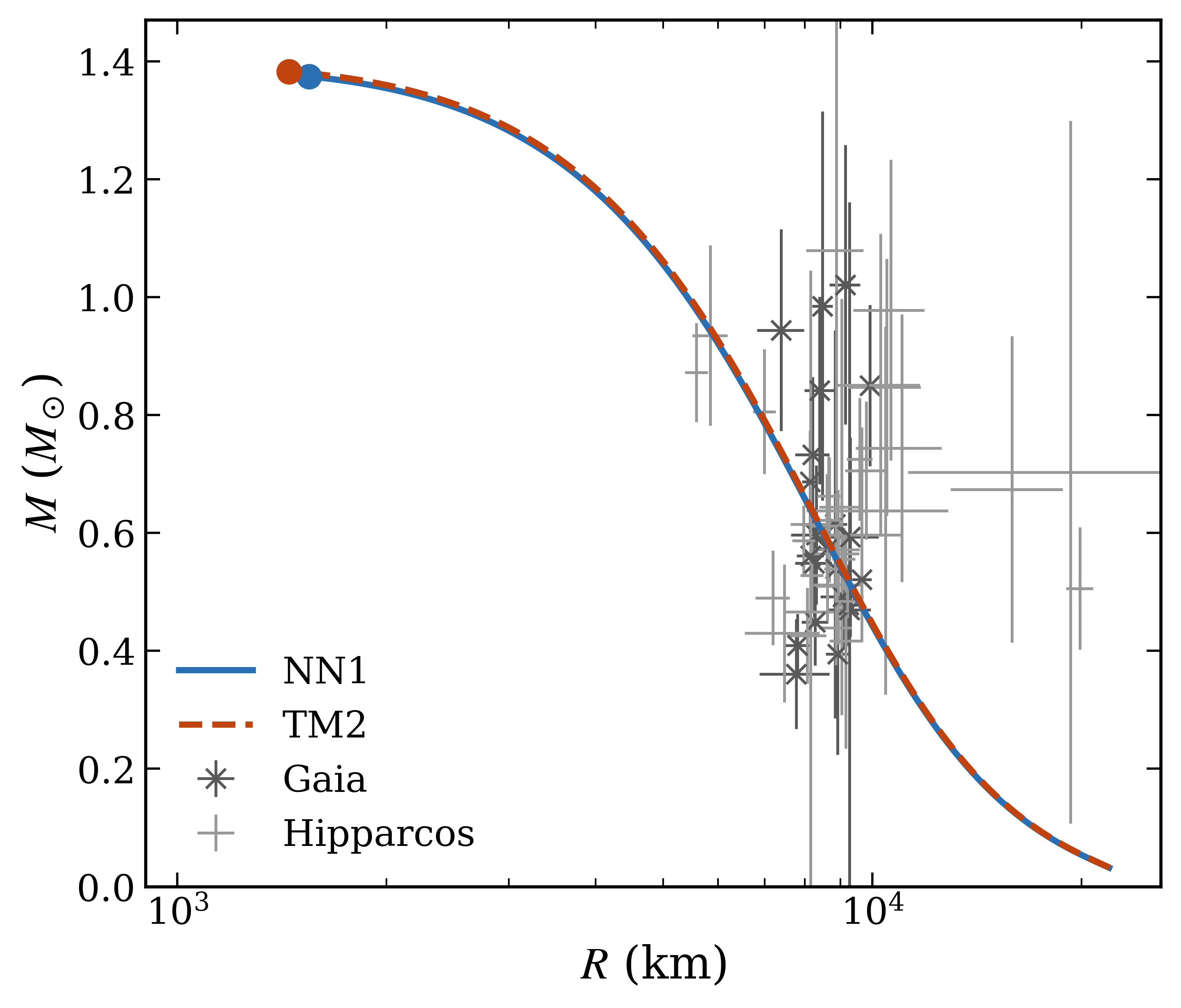}
        \caption{WD branch, compared with the Gaia ($\times$) and Hipparcos ($+$) observational data~\cite{tremblay2016gaia}.}
        \label{fig:mr_tm_wd}
    \end{subfigure}\hfill
    \begin{subfigure}[t]{0.48\textwidth}
        \centering
        \includegraphics[width=\linewidth]{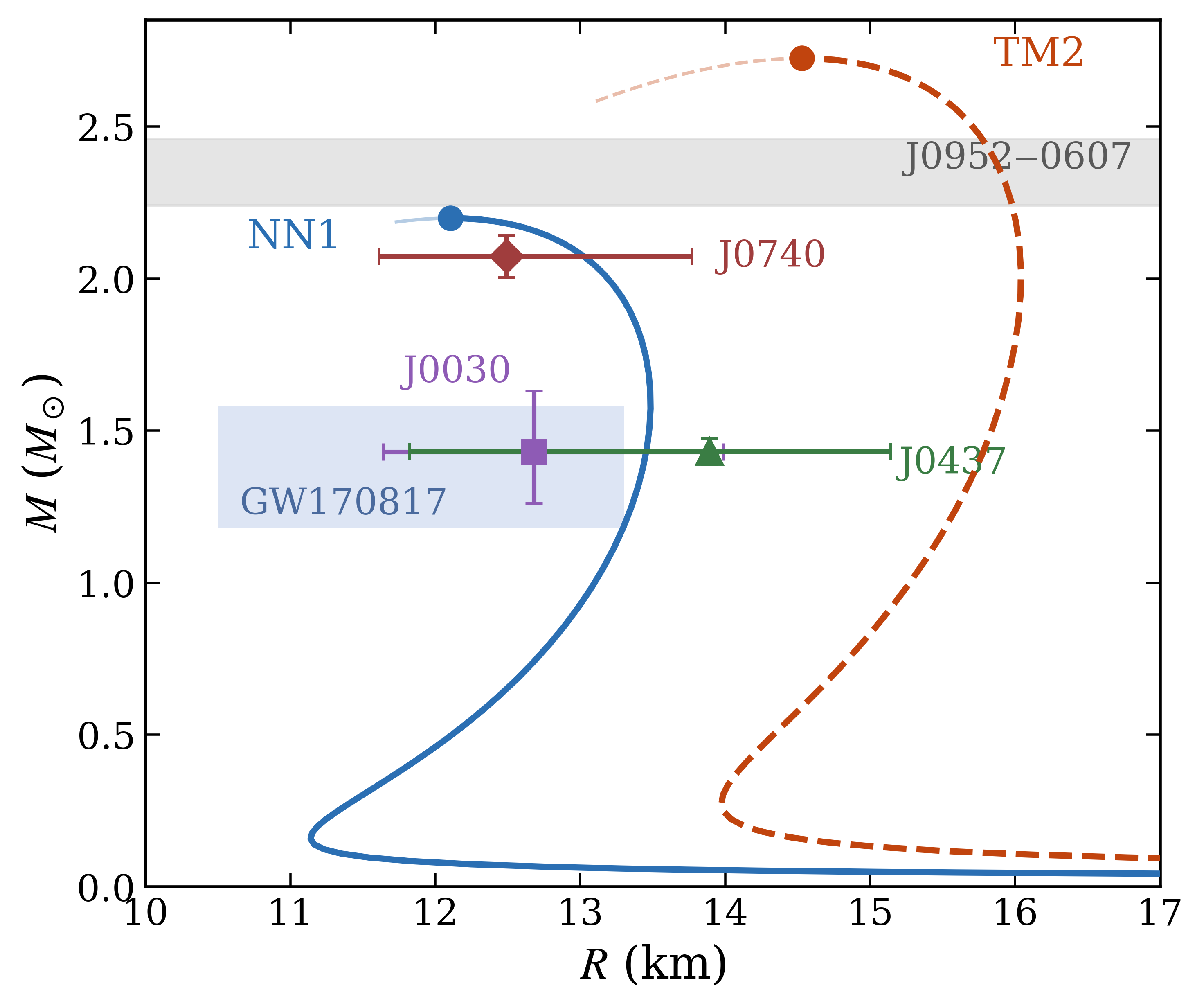}
        \caption{NS branch; filled circles mark the maxima, and faded thin segments denote post-maximum configurations. Overlaid are the 68\% NICER measurements for PSR~J0030+0451~\cite{Kini:2026nicer}, PSR~J0740+6620~\cite{Salmi:2024nicer}, and PSR~J0437-4715~\cite{Miller:2026J0437}, the GW170817 common-EoS component-radius region (90\%)~\cite{LIGOScientific:2018cki}, and the PSR~J0952-0607 mass band ($2.35\pm0.11\,M_\odot$)~\cite{Romani:2022jhd, Romani:2025j0952}, which only the TM2 core EoS reaches.}
        \label{fig:mr_tm_ns}
    \end{subfigure}
    \caption{Mass-radius relations for the $^4$He-seeded sequence comparing the NN1 (solid) and TM2 (dashed) parameter sets, with every junction at its unique equilibrium pressure.}
    \label{fig:mr_tm}
\end{figure*}

Figure~\ref{fig:mr_tm} illustrates the impact of the parameter choice on the M-R relation, comparing NN1 and TM2 for the $^4$He-seeded sequence with all junctions at their unique equilibrium pressures. The WD branches nearly coincide (maximum masses $1.37$ vs $1.38\,M_\odot$), while the stiffer TM2 core EoS yields a heavier and larger NS branch (maximum mass $2.72\,M_\odot$ at $14.5$~km, vs $2.20\,M_\odot$ at $R \simeq 12.1$--$12.2$~km for NN1).

Both sets yield qualitatively similar WD-NS transition curves, but differ in the radii of NS branch,
a direct consequence of the different nuclei properties tabulated above.
In the following discussions we adopt the NN1 parameters, which show better binding-energy agreement than TM2 for the neutron-rich $A=16$ isobars considered here.

\subsection{The complete WD-to-NS mass-radius relation}
\label{subsec:mr_full}

\begin{figure*}[htb]
    \centering
    \begin{subfigure}[t]{0.48\textwidth}
        \centering
        \includegraphics[width=\linewidth]{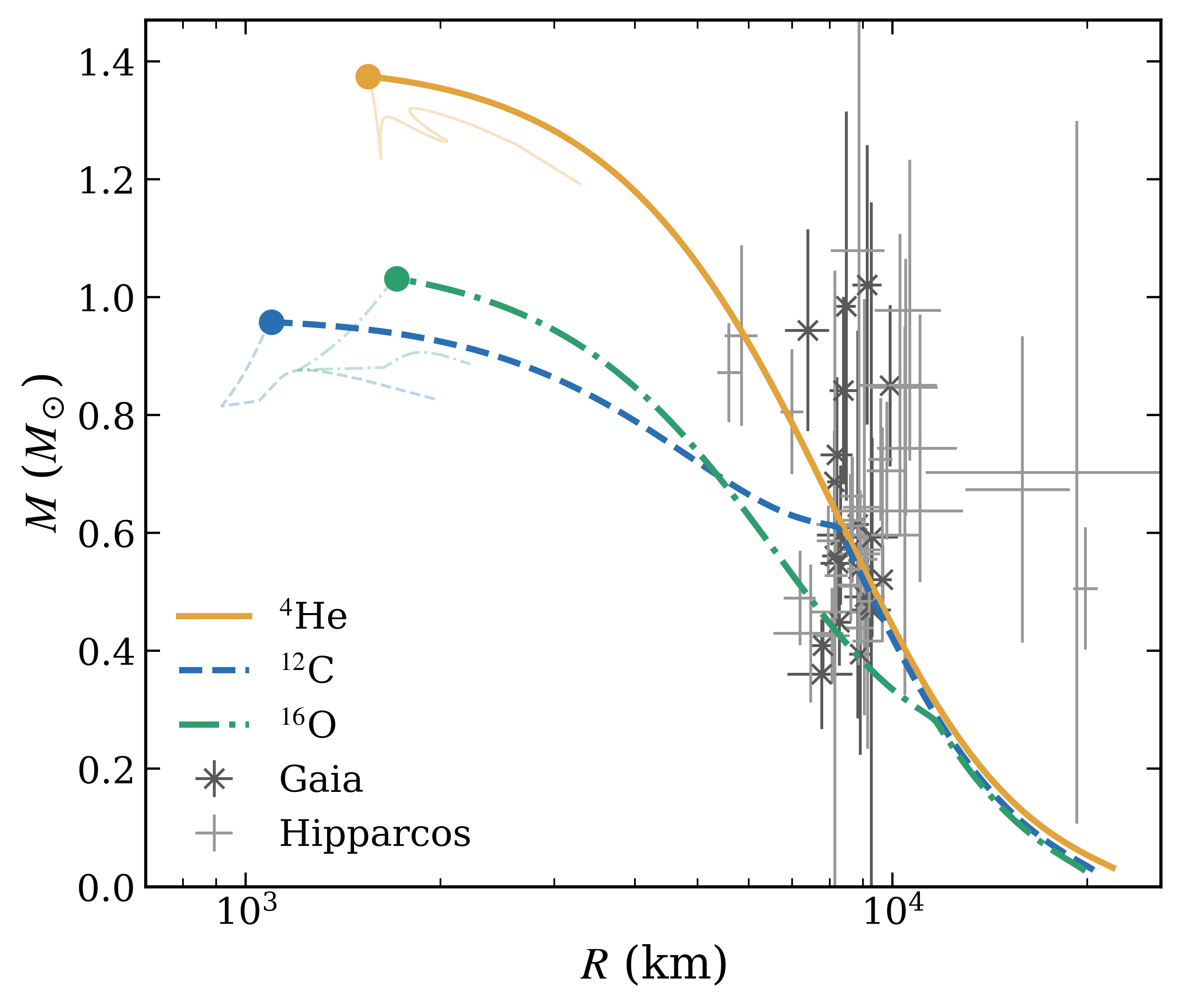}
        \caption{WD branch, compared with the same Gaia ($\times$) and Hipparcos ($+$) observational data as in Fig.~\ref{fig:mr_tm}(a); faded thin segments denote post-maximum configurations. The $^{16}$O-seeded curve contains no pure-$^{16}$O segment: its lowest-pressure stable branch is already the $(9,7)$ ($^{16}$N-like) sub-state.}
        \label{fig:mr_pmu_wd}
    \end{subfigure}\hfill
    \begin{subfigure}[t]{0.48\textwidth}
        \centering
        \includegraphics[width=\linewidth]{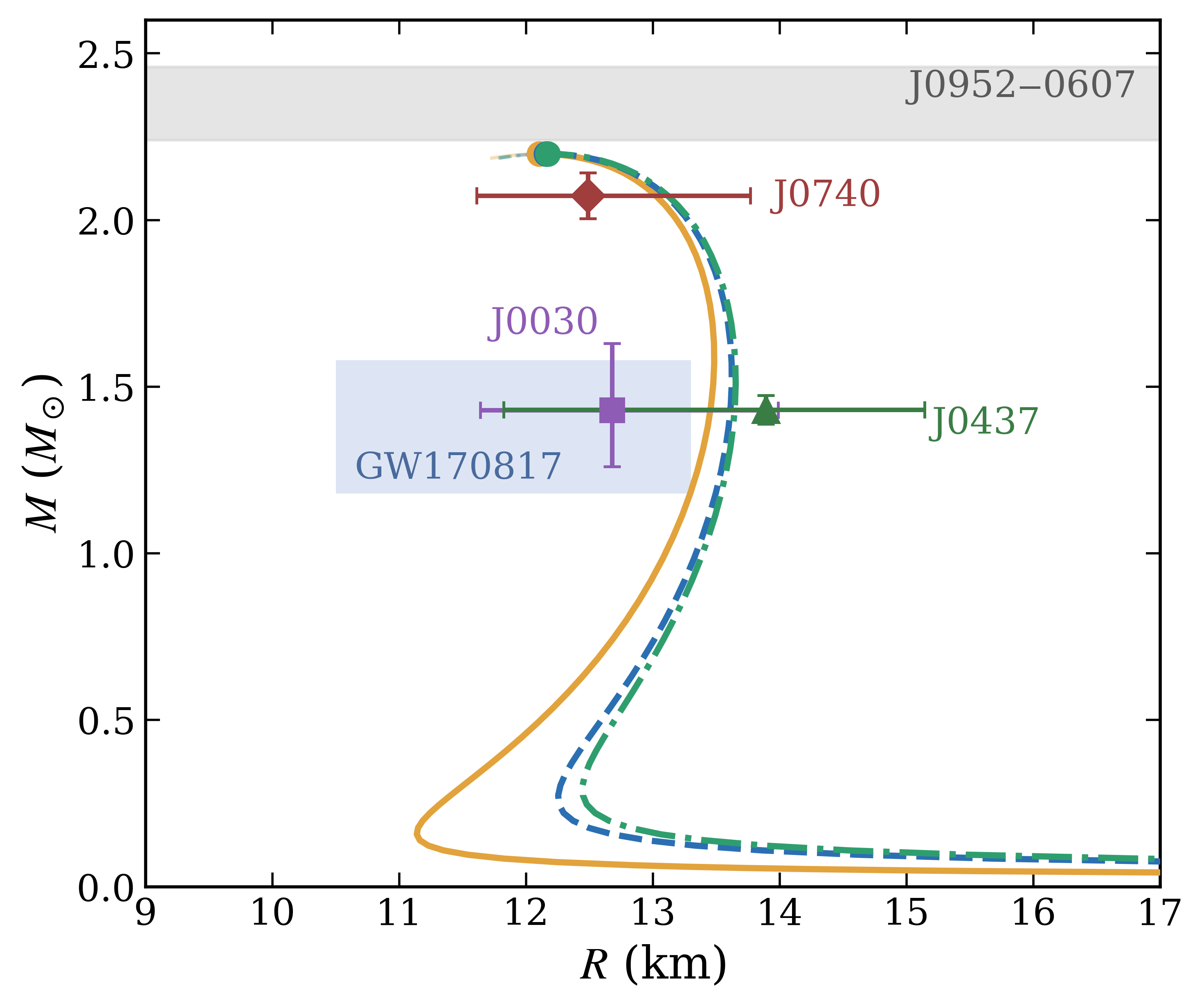}
        \caption{NS branch; the filled circles mark the maxima for the three envelope compositions. Overlaid are the same observational constraints as in Fig.~\ref{fig:mr_tm}(b).}
        \label{fig:mr_pmu_ns}
    \end{subfigure}
    \caption{Mass-radius relations with all junctions at the unique equilibrium pressures, for sequences seeded by $^4$He (solid), $^{12}$C (dashed), and $^{16}$O (dash-dotted).}
    \label{fig:mr_pmu}
\end{figure*}

The principal result of this work is the M-R relation obtained with every junction placed at its unique equilibrium point $p_t$, shown in Fig.~\ref{fig:mr_pmu}. The WD maximum masses are $1.37\,M_\odot$ ($^4$He), $0.96\,M_\odot$ ($^{12}$C-seeded), and $1.03\,M_\odot$ ($^{16}$O-seeded), and the NS branch reaches a maximum mass of $2.20\,M_\odot$ at radii of $12.1$--$12.2$~km for the three envelope compositions, since the core EoS is common.

Using the NN1 parameters, Fig.~\ref{fig:mr_pmu} presents the stable WD and NS branches of the M-R relations for the $^4$He-, $^{12}$C-, and $^{16}$O-seeded sequences. The elemental composition of the progenitor dominates the WD branch, while the NS branch is controlled by the common core EoS: the NS maximum mass is composition independent, and the envelope composition survives as a radius spread that grows toward lower masses (${\sim}0.1$~km at $2.0\,M_\odot$, ${\sim}0.2$~km at $1.4\,M_\odot$, ${\sim}0.3$~km at $1.0\,M_\odot$, with $^4$He giving the smallest radii; cf.\ Sec.~\ref{subsec:mr_ns}).

\begin{figure*}[htb]
    \centering
    \includegraphics[width=0.85\textwidth]{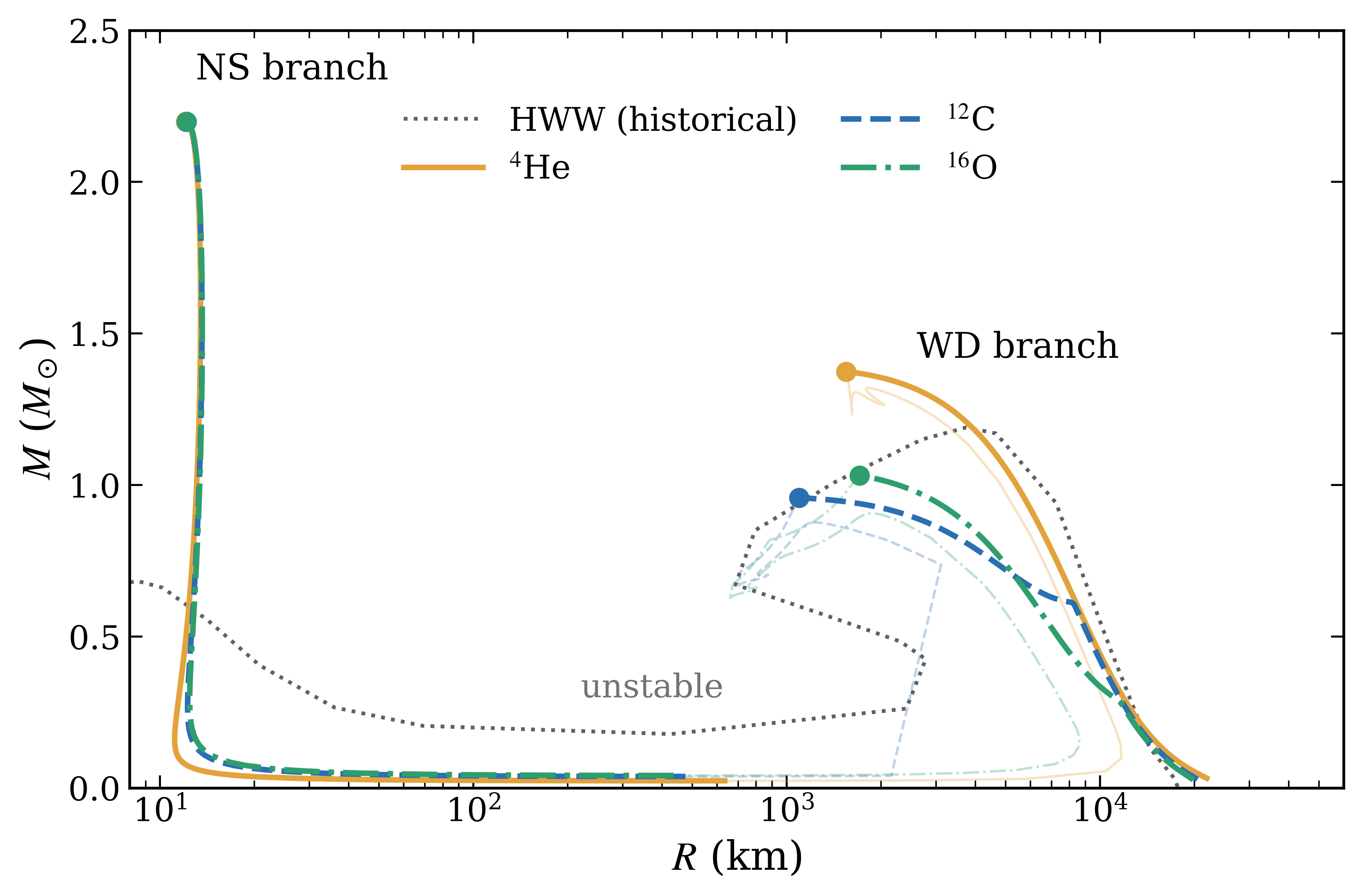}
    \caption{Complete mass-radius relations of the unified EoSs for the $^4$He- (solid), $^{12}$C- (dashed), and $^{16}$O-seeded (dash-dotted) sequences, with the historical nonrotating Harrison--Wakano--Wheeler (HWW) sequence shown as a thin gray dotted curve~\cite{Harrison:1965gtc, Hartle:1968hww}. Thick colored segments satisfy $dM/dn_c > 0$, faded thin segments have $dM/dn_c < 0$, and filled circles mark the WD and NS maxima.}
    \label{fig:mr_full}
\end{figure*}

Fig.~\ref{fig:mr_full} assembles the two branches satisfying the necessary turning-point condition and the intermediate unstable interval into the complete equilibrium sequence, the analogue---for our restricted fixed-$A$ EoSs---of the classical cold-equilibrium family first mapped from WD to NS densities by Harrison, Thorne, Wakano, and Wheeler~\cite{Harrison:1965gtc}. The overlaid HWW sequence uses the nonrotating configurations recalculated and tabulated by Hartle and Thorne~\cite{Hartle:1968hww}. It shows that the classical and present sequences share the WD--unstable--NS topology while differing quantitatively because their composition constraints and high-density EoSs are not the same. Along a one-parameter sequence of cold equilibria, radial stability changes at extrema of $M(n_c)$~\cite{Harrison:1965gtc, Shapiro:1983du}, so the branch between the WD maximum and the minimum of the core-bearing branch is radially unstable. It should be stressed that this turning-point rule is an empirical criterion inherited from earlier studies of cold-equilibrium families~\cite{Chandrasekhar:1964rad, Meltzer:1966, Sorkin:1981}, and provides a necessary but not a sufficient condition for radial stability; for configurations containing a sharp interface the appearance of a new core is further constrained by the density jump at the transition~\cite{Seidov:1971, Alford:2013}. In cold catalyzed matter this unstable interval opens near the neutron-drip density $\rho_{\rm drip} \simeq 4.3\times10^{11}~\mathrm{g\,cm^{-3}}$~\cite{Baym:1971pw}, and representative unified EoSs place the NS minimum at $M_{\rm min} \simeq 0.088$--$0.094\,M_\odot$ with $R \simeq 220$--$270$~km (FPS and SLy; the older BPS model gives $0.0925\,M_\odot$ at $R \simeq 164$~km)~\cite{Baym:1971pw, Haensel:2002cia}. These equilibrium limits are distinct from astrophysical formation limits: ultra-stripped core-collapse calculations yield remnant gravitational masses no smaller than ${\sim}1.17\,M_\odot$~\cite{Suwa:2018uni}, and accretion-induced collapse of massive WDs~\cite{Nomoto:1991aic} typically leaves ${\sim}1.2$--$1.3\,M_\odot$ remnants.

In the present construction the low-density phase is restricted to fixed-$A$, non-catalyzed Wigner-Seitz matter, so the quantitative landmarks differ from the catalyzed benchmarks: the minimum of the core-bearing branch occurs at $M \simeq 0.02$--$0.04\,M_\odot$ with $R \sim 10^2$--$2\times10^3$~km, reflecting the extended fixed-$A$ envelope carried by low-mass configurations rather than a revision of the canonical catalyzed minimum. We quote this value as a feature of the restricted EoS; establishing it as a dynamically stable minimum would additionally require a radial-oscillation analysis with the frozen-composition adiabatic response~\cite{Chanmugam:1977, Haensel:2002adi} and explicit junction conditions (rapid versus slow phase conversion) at the Maxwell density discontinuities~\cite{Colpi:1993qc, Haensel:1989, Pereira:2018}. With these caveats, the complete curves of Fig.~\ref{fig:mr_full} provide the equilibrium baseline on which accretion-induced-collapse trajectories and NS-WD merger simulations---including their decihertz gravitational-wave signatures~\cite{Kang:2024wdns}---can be built, although the cold equilibrium sequence itself should not be read as a dynamical collapse path.

\subsection{The M-R relation for WD}
\label{subsec:mr_wd}

Fig.~\ref{fig:mr_pmu}(a) compares the WD branch with the Gaia and Hipparcos observational data~\cite{tremblay2016gaia}. With the full low-density envelope included, the model sequences pass through the bulk of the observed population: at $M \simeq 0.6\,M_\odot$ the predicted radii are $\sim 8\times10^3$~km, consistent with the data cloud.

For the $^4$He WD, the maximum mass is approximately $1.4\,M_\odot$ with a radius of $\sim 1.5\times10^3$~km at that maximum-mass configuration, consistent with the bulk of the observed WD population. 
For the $^{12}$C-seeded WD, the maximum mass is approximately $1.0\,M_\odot$ with a radius of $\sim 1.1\times10^3$~km at that maximum-mass configuration. Near this maximum, most of the star lies on the $(7,5)$ branch, where only five electrons are present per 12 baryons ($Y_e=5/12$, or $A/Z=2.4$), rather than the $Y_e=1/2$ of neutral $^{12}$C. Because the Chandrasekhar mass scales as $M_{\rm max}\propto Y_e^2$, this reduction gives $1.44\,M_\odot[(5/12)/(1/2)]^2\simeq1.0\,M_\odot$.

These maxima lie below the classical Chandrasekhar limit (${\sim}1.44\,M_\odot$) is because that our framework couples the WD structure directly to the microscopic energetics of the carbon isobars: neutronization proceeds through the sub-states of Sec.~\ref{sec:transition}, so the M-R relation probes the in-medium $^{12}\mathrm{C}\to{}^{12}\mathrm{B}$ electron-capture threshold rather than assuming a fixed electron fraction.
With the NN1 description the zero-pressure energy splitting between the $(7,5)$ and $(6,6)$ cells is $0.025$~MeV per-baryon, so capture sets in early along the sequence, the electron fraction $Y_e$ is reduced, and the maximum mass follows $M_{\rm max}\propto Y_e^2$ down to ${\sim}1.0\,M_\odot$.
The size of this departure is set by how accurately the relevant isobars are described and can be sharpened by improving the calculation of these nuclei---for example through in-medium corrections beyond the present mean-field treatment, whose vacuum reference is the evaluated $^{12}\mathrm{B}$--$^{12}\mathrm{C}$ mass difference of $1.114$~MeV per-baryon~\cite{Wang:2021AME}.

No $^{16}$O WD branch appears in our calculation: the $(8,8)$ sub-state never minimizes the Gibbs free energy per-baryon, so the restricted $A=16$ path begins on the $(9,7)$ ($^{16}$N-like) branch.
This behavior is governed by the balance between the binding of $^{16}$O relative to its neutron-richer isobars and the electron energetics of the extra electron per-two-baryon (the $(8,8)$ cell carries $Y_e=1/2$ rather than $7/16$): with the NN1 description $^{16}$O is more bound than $^{16}$N by ${\sim}0.10$~MeV per-nucleon, whose vacuum reference is the evaluated $0.61$~MeV per-nucleon associated with the doubly closed shells of $^{16}$O~\cite{Epelbaum:2013paa,Halcrow:2020skg}.
As for carbon, the location of the oxygen sequence is thus a sensitive probe of the in-medium isobar energetics, and improving the calculation of the corresponding nuclei may either restore an $^{16}$O branch or confirm the present ordering.

\subsection{The M-R relation for NS}
\label{subsec:mr_ns}

We next consider the M-R relation of NS. In our model, the low-density envelope (the outer-crust region) of a NS is naturally provided by the lighter-element WD matter at the high-density end of the EoS.
The elemental composition of this envelope cannot be fixed. We illustrate its effect on the predicted NS structure by varying it across $^4$He, $^{12}$C, and $^{16}$O.
Fig.~\ref{fig:mr_pmu}(b) shows the resulting NS M-R relations with every junction at its unique equilibrium pressure, overlaid with the current observational constraints: illustrative crosses showing the reported marginal 68\% mass and radius intervals for PSR~J0030+0451 ($M=1.43^{+0.20}_{-0.17}\,M_\odot$, $R=12.68^{+1.31}_{-1.04}$~km~\cite{Kini:2026nicer}), PSR~J0740+6620 ($2.073\pm0.069\,M_\odot$, $12.49^{+1.28}_{-0.88}$~km~\cite{Salmi:2024nicer}), and PSR~J0437-4715 ($1.431\pm0.043\,M_\odot$, $13.89^{+1.25}_{-2.07}$~km~\cite{Miller:2026J0437}); the GW170817 common-EoS component-radius region $R = 11.9\pm1.4$~km (90\%, component masses $1.18$--$1.58\,M_\odot$)~\cite{LIGOScientific:2018cki}; and the PSR~J0952-0607 mass band $2.35\pm0.11\,M_\odot$~\cite{Romani:2022jhd, Romani:2025j0952}. The model curves overlap the displayed marginal intervals of J0030 and J0740 and the upper part of the GW170817 region, and are consistent with the broad J0437 interval.

The predicted NS maximum mass, $2.20\,M_\odot$, is consistent with the measured mass of PSR~J0740+6620 ($M = 2.08 \pm 0.07\,M_\odot$~\cite{Fonseca:2021wxt}).
At the pulsar-timing mass of PSR~J0437-4715, $M = 1.418\,M_\odot$~\cite{Reardon:2024J0437}, our sequences give $R = 13.4$--$13.6$~km. This lies within the 68\% credible range of the 2026 NICER analysis incorporating modulated nonthermal emission, $11.82$--$15.14$~km~\cite{Miller:2026J0437}, though above the earlier 68\% interval $R = 11.36^{+0.95}_{-0.63}$~km~\cite{Choudhury:2024J0437}; the two analyses differ in their emission modeling and should not be averaged.

The retained low-density envelope prescription produces a percent-level composition dependence, approximately $0.2$~km at $1.4\,M_\odot$.
Helium gives the smallest radius among the three model sequences, while $^{16}$O gives the largest, and seed compositions and possible accretion histories can be distinguished if a precise radius measurement can be made for a intermediate/low-mass NS.

\section{Discussion and outlook}
\label{sec:concl}

We have constructed a unified EoS spanning the full density range from WD to NS matter within a single RMF framework.
WD matter is described via the Wigner-Seitz cell model, in which a self-consistently solved nucleus is surrounded by a uniform electron gas and the full electromagnetic interaction is treated via Maxwell equations.
NS matter is computed in the uniform NM approximation using the same Walecka-type Lagrangian.
The neutronization path connecting the two regimes is tracked through the beta-equilibrium. The preferred sub-state $(N_n, N_p)$ is selected at each pressure as the one minimizing the Gibbs free energy per-baryon, and adjacent sub-states are joined at the crossing points of their $p(\mu_B)$ curves, where mechanical and chemical equilibrium hold simultaneously and the baryon density jumps discontinuously.
Taking $^4$He, $^{12}$C, and nominal $^{16}$O initial seeds for the restricted sequences, we solve the TOV equations to obtain M-R relations across the WD-to-NS transition. We integrate the TOV equations with an adaptive Runge-Kutta scheme on the tabulated EoS with logarithmic interpolation, using the central density as the sequence parameter and the surface criterion $p_{\rm surf} = 10^{-17}~\mathrm{MeV\,fm^{-3}}$ discussed in Sec.~\ref{sec:struc}; central densities falling inside a Maxwell density gap are excluded, and each first-order interface is represented by the two coexistence points at equal pressure. 

Several key findings emerge in this work.
Our NN1 parameter set with physical nucleon masses ($m_n \neq m_p$) reproduces the binding energies of neutron-rich $A=16$ isobars significantly better than the standard TM2 set, confirming its suitability for the neutron-rich environment of the WD-to-NS transition.
For the WD branch, the maximum masses are ${\sim}1.4\,M_\odot$ ($^4$He) and ${\sim}1.0\,M_\odot$ ($^{12}$C), while no $^{16}$O ground state appears because the $(8,8)$ sub-state never minimizes the Gibbs free energy per-baryon. Both the reduced carbon maximum and the absence of an $^{16}$O branch follow from coupling the WD structure to the in-medium energetics of the corresponding isobars. Therefore, they are sensitive probes of that nuclear input, and improved calculations of these isotopes---which may themselves yield mass-radius relations of this character---are a natural direction for further study.
On the NS branch, the transition placement is uniquely fixed by the equilibrium conditions, leaving the elemental composition of the envelope and the core EoS as the only physical degrees of freedom.
A feature of our approach is the unified treatment of the outer layers by retaining the low-density WD-derived envelope as part of the NS EoS rather than truncating at a fixed surface density; the resulting composition dependence of the radii is at the percent level and lies within current observational uncertainties.
Helium gives the smallest radius among the three model sequences and $^{16}$O the largest. Hence, a precise radius measurement of an intermediate- or low-mass NS could therefore distinguish the seed composition, and hence the accretion history of the star.

The unified EoS developed here is a natural starting point for studying WD-NS binaries and their mergers.
WD-NS inspirals emit gravitational waves in the decihertz band ($\sim 0.01$--$1$~Hz), where DECIGO is projected to detect tens of thousands of events per-year, with BBO offering comparable sensitivity~\cite{Kang:2024wdns}, and their merger remnants have been proposed as progenitors of peculiar long-duration gamma-ray bursts, including GRB~211211A~\cite{Liu:2025grb} and GRB~230307A~\cite{Du:2024grb230307, Wang:2024grb230307}.
A unified EoS that covers both WD and NS densities is a prerequisite input for hydrodynamic merger simulations and r-process nucleosynthesis calculations, once extended to finite temperature and out-of-equilibrium composition.
The most natural next calculation is the tidal Love number and deformability of the low-mass NS branch, where the seed-composition dependence of the outer layers enters through the $R^5$ scaling of $\Lambda$. The unified EoS constructed here is also, with appropriate finite-temperature and reaction-network extensions, a starting point for accretion-induced-collapse studies, although the present cold equilibrium sequence should not itself be read as a collapse trajectory.
The radial stability of the present sequences has been assessed only through the empirical turning-point criterion adopted from earlier studies of cold-equilibrium families. A more rigorous study should therefore be carried out on the basis of a realistic radial-pulsation calculation---solving the radial oscillation equations with the frozen-composition adiabatic index and explicit junction conditions for rapid and slow phase conversion at each first-order interface~\cite{Haensel:1989, Pereira:2018}---in order to establish which segments of the complete sequence are genuinely stable.
It is also promising to connect our model to WD progenitor evolution. By using the relation between progenitor mass and WD core composition~\cite{1996ApJ...460..489R, Herwig:2005zz, 2017A&A...597A..67A} together with asteroseismological constraints on WD internal structure~\cite{1988IAUS..123..305W, Winget:2008iu, 2017EPJWC.15201011K}, one could trace the full stellar evolution path from main sequence through WD to NS, as well as the neutrino (and/or dark matter) emission.

The NS crust---spanning from the crystallized outer crust through the pasta-phase inner crust to the liquid core~\cite{Chamel:2008ca}---governs several observational phenomena, including pulsar glitches~\cite{Anderson:1975zze}, NS cooling~\cite{Yakovlev:2004iq}, and the tidal deformability measured in binary inspirals.
Because our model follows a single restricted neutronization path inherited from the WD composition, with every junction fixed by chemical and mechanical equilibrium, it is convenient to study these quantities within one framework. A self-consistent treatment of the inner crust between the Wigner-Seitz branches and uniform matter (clusters with dripped neutrons, Coulomb and surface energies, pasta phases) remains the principal missing ingredient.
The low-density end of our neutronization sequence is analogous to the conventional Baym--Pethick--Sutherland (BPS) outer crust~\cite{Baym:1971pw}, with the difference that our treatment retains the light-element composition inherited from the WD within a single continuous crust-to-core EoS, whereas the cold-catalyzed BPS crust follows the energetically preferred iron-peak-to-neutron-drip sequence. A systematic comparison of the two descriptions is left to future work.
More realistic WD modeling---including non-spherical geometry, spin distortion, tidal deformation~\cite{McNeill:2019rct}, and thermal and crystallization effects in cold WDs~\cite{Perot:2022zwy}---would further refine the outer-envelope boundary condition and extend the model's predictive reach.
Replacing the present Maxwell construction with a Gibbs mixed-phase construction~\cite{Glendenning:2000}, extending the single-element crust into a full multi-nuclide BPS-type cold-catalyzed optimization~\cite{Baym:1971pw}, or an inner-crust model that self-consistently includes nuclear pasta phases~\cite{Ravenhall:1983uh, Chamel:2008ca} represents a natural refinement of the transition region.
Finally, incorporating exotic degrees of freedom---strange quark matter in NS cores~\cite{Alcock:1986hz,Alcock:1988re,Madsen:1998uh}, quark matter inside WD cores~\cite{Benvenuto:2005xs}, or hybrid and quark star configurations~\cite{Weber:2004kj}---represents a natural extension of the present framework toward more general compact-star phenomenology.

\section*{ACKNOWLEDGMENTS}

The authors would like to thank Dr. Ling-Jun Guo to his helpful discussions on the algorithm of the Wigner-Seitz cell solver.

The work of Y. M. is supported by Jiangsu Funding Program for Excellent Postdoctoral Talent under Grant Number 2025ZB516.
Y.~L. M. is supported in part by the National Science Foundation of China (NSFC) under Grant No. 12547104, the National Key R\&D Program of China under Grant No. 2021YFC2202900 and Gusu Talent Innovation Program under Grant No. ZXL2024363.
Y.~L. W. was supported in part by the National Science Foundation of China (NSFC) under Grants No. 12547104 (special fund to the center for quanta-to-cosmos theoretical physics), No. 11821505, the National Key Research and Development Program of China under Grant No. 2020YFC2201501, and the Strategic Priority Research Program of the Chinese Academy of Sciences.

\section*{DATA AVAILABILITY}

The program implementing the framework developed in this work---the RMF Wigner-Seitz cell solver, the neutronization and Maxwell-construction routines, and the TOV integration used to obtain the mass-radius relations---is openly available at \url{https://github.com/AaronMahn/Uni-WD-NS}, together with the input parameter sets and the scripts that reproduce all EoSs, tables and figures presented here.

\bibliography{ref}

\end{document}